\documentclass[aps,prl,twocolumn,showpacs,floatfix]{revtex4}
\usepackage{graphicx}
\usepackage{amsmath}
\usepackage{amsfonts}
\usepackage{amssymb}
\usepackage{bm}
\begin{document}

\title{Symmetry consideration in identifying network structures}
\author{Jiao Wang$^{1,3}$ and C.-H. Lai$^{2,3}$}
\affiliation{$^{1}$Temasek Laboratories, $^{2}$Department of
Physics and
\\$^{3}$Beijing-Hong Kong-Singapore Joint Center for Nonlinear and Complex Systems (Singapore)
\\National University of Singapore, 117542, Singapore}

\begin{abstract}
The topological information of a network can be retrieved
equivalently from its complement consisting of the same nodes but
complementary edges. Hence the partition of a network into certain
substructures based on given criteria should be the same as that
of its complement based on the {\it equivalent} criteria if the
topological information is considered exclusively. This symmetry
of partitioning between a network and its complement is due to the
equivalence of their topological information and hence should be
respected regardless of the detailed characteristics of the
substructures considered. In this work we suggest this symmetry
consideration as a general guideline and propose a symmetric
community detection scheme to show its implications. Our method
has no resolution limit and can be used to detect hierarchical
community structures at different levels. Our study also suggests
that the community structure is unlikely a result of random
fluctuations in large networks.
\end{abstract}

\pacs{89.75.Hc, 89.75.Fb, 05.45.-a} \maketitle

In the last decade complex networks have been extensively studied
with the aim to reveal and understand their structures at various
scales \cite{rev}.  Besides the general statistical properties
such as small-world \cite{sw} and scale-free \cite{sf} properties,
the significance of some common structural features at the
mesoscopic level has also been realized. The mesoscopic structures
having received intensive studies include communities \cite{cmrv}
and similar groups \cite{Rosv,NL,WL}, i.e. node sets whose
components have similar connection patterns. These mesoscopic
structures are of scientific interest because they may have a
close relation to certain behavioral or functional units of the
system \cite{bisoc}, and meanwhile they provide an ideal basis for
reduction or coarse-graining of networks \cite{Rosv,redc}, which
could be particularly useful in dealing with networks of huge size
as often encountered nowadays. Furthermore, these substructures
also have important implications for various dynamical processes
over the networks \cite{dynm,hier}.

However, in spite of the efforts and fast progress made in this
field, the detection of these substructures still remains
challenging. (Here we restrict ourselves to networks consisting of
these mesoscopic structures exclusively, such that the problem of
detection is equivalent to that of partitioning.) One conceptual
difficulty is the ambiguity in the definition of these
substructures \cite{cmrv}, and the question of what
characterizations are essential to them has not been thoroughly
understood yet. In this Letter we suggest a symmetry that should
be taken into account in the definition of network structures. It
does not address the details of individual substructures and their
characterizations, but is a property of networks. It provides a
consistency criterion with which the network structures can be
specified more precisely. The detection of network structures can
then be improved as a result.

This symmetry originates from the dual nature of connection states
in networks. Consider a network of $N$ nodes whose connection
topology is encoded in the adjacency matrix $A$ with $A_{ij}=1$ if
node $i$ and $j$ are connected and $A_{ij}=0$ otherwise.
Obviously, the topological information contained in $A$ is
completely equivalent to that in its complement $\overline A$
related to $A$ via the one-to-one map $\overline A
_{ij}=1-A_{ij}$. (The network corresponding to $\overline A$ is
referred to as the complement of the network corresponding to
$A$.) Due to this equivalence, it is natural to expect that any
structure recognized in network $A$ based on certain
characterizations should be recognized in network $\overline A$
based on the same or equivalent characterizations. By way of
analogy, this is similar to recognizing a face in a photo; given
its features it can be done in the negative film equivalently.
Assuming this equivalence, it provides an approach to check if the
characterizations used for defining a structure are consistent.
Only those characterizations (and their equivalent) based on which
we can recognize the same structures in a network and its
complement are regarded as consistent and acceptable. We suggest
this symmetry principle should be adopted as a necessary condition
in defining the network structures.

This symmetry consideration has not been adopted as a general
guideline in most investigations. In a recent study \cite{WL} a
symmetric definition of similar groups is proposed. It has been
found that indeed the symmetric definition can overcome some
difficulties encountered with the asymmetric definition \cite{NL}.
Moreover, the symmetric definition can be extended to the
connection information weighted networks, resulting in a new
perspective to see the role of weights in the problem \cite{WL}.
It is interesting to notice that a symmetric definition of the
community in a general spectral detection algorithm has also been
found to outperform the asymmetric definitions \cite{sptm}.

\begin{figure}
\includegraphics[width=.8\columnwidth,clip]{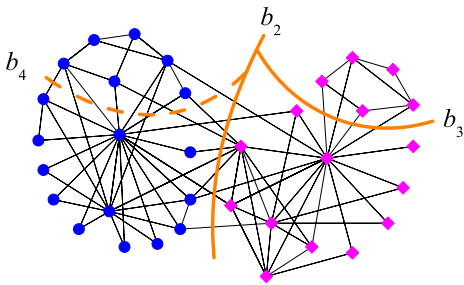}
\vspace{-.3cm} \caption{(color online) The optimal community
partitioning of the karate network \cite{karate} corresponding to
the maximum value of the symmetric quality function with given
number of modules $C=2, 3$ and $4$ represented by the partition
lines $b_2$ ($C=2$), $b_2$ and $b_3$ ($C=3$), $b_2$, $b_3$ and
$b_4$ ($C=4$), respectively. The largest symmetric modularity
corresponds to the partition given by $b_2$ and $b_3$ (see Fig. 2
(b)).}
\end{figure}

To demonstrate the power of this symmetry guideline, in the
following we apply it to the community detection problem by
constructing a symmetric quality function of partition. An
asymmetric version, which has been so far the most  popular
\cite{cmrv}, is that suggested by Newman and Girvan \cite{NG}. For
a given partition $\pi^C$ of network $A$ that contains $C$
modules, it reads
\begin{eqnarray}
q_{NG}(A,\pi^C)=
\sum_{\alpha=1}^C[\frac{l_\alpha}{L}-(\frac{d_\alpha}{2L})^2],
\label{eq-qNG}
\end{eqnarray}
where $L$ is the total number of links in the network, $d_\alpha$
is the total degree of nodes in module $\alpha$ and $l_\alpha$ is
the number of internal links of module $\alpha$. The summand
represents how much the fraction of links inside a module is more
than what is being expected in the {\it null model} of $A$, i.e.
random networks sharing the same nodes and the same degree sequence.
For convenience let us denote by $M^C_{NG}$ the maximum of
$q_{NG}$ over all the possible partitions containing $C$ modules;
then the {\it modularity}, ${M}_{NG}$, is defined as the maximum
value of $M^C_{NG}$ over all allowed $C$ values, and the
corresponding partition is regarded to be the optimal community
partition of network $A$ \cite{NG}.

The concept of modularity is an important contribution to the
definition and detection of communities in networks \cite{cmrv}.
The modularity maximization has itself been developed into a
popular method and many algorithms have also been developed for
this purpose. Some issues however remain to be addressed. One is
that the modularity based methods have a resolution limit
$\sim\sqrt L$ preventing them from identifying communities smaller
than this limit \cite{reslmt}; another is that modularity may
attain fairly large values when being applied to partitioning
random networks \cite{bigflt}, making its meaning elusive
\cite{cmrv}.

\begin{figure}
\includegraphics[width=.9\columnwidth,clip]{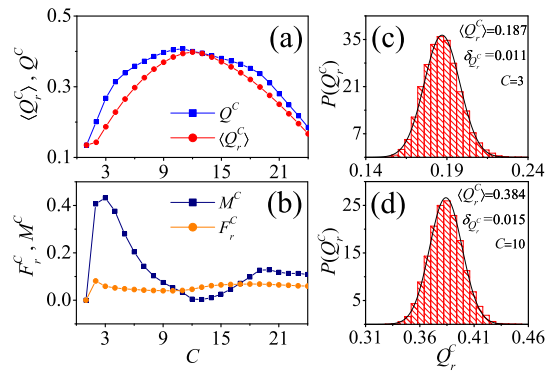}
\vspace{-.1cm} \caption{(color online) The analysis of the karate
network \cite{karate}. The comparison of $Q^C$ and $\langle
Q_r^C\rangle$ (a) and that of $M^C$ and $F_r^C$ (b) where $\langle
Q_r^C\rangle$ and $F_r^C$ are evaluated over $10^4$ random
networks with the same degree sequence. (c) and (d) show the
distribution of $Q_r^C$ for $C=3$ and $C=10$ respectively. Solid
curves are Gaussian with the same averages and deviations. }
\end{figure}

We consider here instead a symmetric quality function of
partition:
\begin{eqnarray}
q(A,\pi^C)=\frac{1}{N}\sum_{\alpha=1}^C[\frac{d_\alpha^{in}}
{N_\alpha}-\frac{d_\alpha^{out}}{N-N_\alpha}]. \label{eq-q}
\end{eqnarray}
Here $N_\alpha$ is the number of nodes in module $\alpha$,
$d_\alpha^{in}$ ($d_\alpha^{out}$) is the total degree of nodes in
module $\alpha$ corresponding to their connections to themselves
(other modules). The summand reflects the difference between the
average edges a node in a module and a node outside can have to
connect to the nodes in that module. For the partition $\pi^1$
that all $N$ nodes are assigned into a single module ($C=1$), it
can be naturally extended to $q(A,\pi^1)\equiv d/{N^2}$ ($d$ is
the total degree of all nodes). Apparently, $q$ thus defined is
symmetric; i.e. $q(A,\pi^C)=-q(\overline A,\pi^C)$.

Of all the possible partitions that have $C$ modules the one,
denoted by ${\tilde \pi}^C$, that generates the maximum $q$ value
is regarded to be the optimal community partition {\it given} $C$.
This is in agreement with our expectations for a good community
partition. As $q(A,{\tilde \pi}^C)=-q(\overline A,\tilde \pi^C)$,
it suggests that to find the optimal partition with $C$ modules in
$A$ by maximizing $q$ can be equivalently done by minimizing $q$
in $\overline A$. For this reason $q$ is more consistent. We
denote by $Q^C\equiv q(A,{\tilde \pi}^C)$ for the sake of
convenience.

As an example Fig. 1 shows the optimal partition ${\tilde \pi}^C$
of the karate network \cite{karate} with $C=2,3$ and $4$. Indeed
they are consistent with our intuition of communities. In all
other networks we have investigated this is always the case.
Numerically we employ an accurate and very efficient fusion
algorithm (AdClust) \cite{AdClust} with a slight modification.
Initially each node consists of a module. At each fusion step
followed there are two operations. First, for each node one finds
the target module which moving the node into may generate a
maximum positive increase of $q$ value. If this is successful then
the node is moved into that module. After all the nodes are
considered (one by one and with a random order) this process is
repeated with a new random order until all nodes are stable. Next,
for all possible module pairs one finds the one whose merger may
lead to the maximum increase (or minimum decrease) of the $q$
value and then combine them. These two operations are repeated
until all the modules evolve into one. During this process we can
obtain a series of partitions of different numbers of modules, and
they are regarded to be good approximations of the optimal
partition ${\tilde \pi}^C$. Careful studies have shown that
different node orders taken in the first operations may lead to
different partition results. For this reason $10^3\sim 10^4$
`random realizations' are performed in our calculations and the
largest $q$ values and the corresponding partitions are chosen to
be the final approximations of $Q^C$ and ${\tilde \pi}^C$. We have
also checked the results obtained in this way with the stimulated
annealing algorithm \cite{SA} and found that they cannot be
improved any further.

Next, let us find out, among all the partitions with different
number of modules $\{{\tilde \pi}^C, C=1,2,...\}$, which one could
be the most relevant. For this purpose we consider the null model,
i.e. random networks that share the same degree sequence with the
network considered, and define the symmetric modularity for a
given $C$ as
\begin{eqnarray}
M^C=\frac{Q^C-\langle Q_r^C\rangle}{\langle Q_r^C\rangle}.
\label{eq-MC}
\end{eqnarray}
Here $Q_r^C$ is the maximum $q$ value for the optimal partition of
a random network of null model, and ${\langle Q_r^C\rangle}$ is
the corresponding average over all such networks. $M^C$ measures
how much more modular the communities found in the original
network are as compared with those found in the corresponding
random networks. If the communities are seen as certain ordered
structures, then $M^C$ also reflects how orderly the communities
found are as compared with their counterparts arising out of pure
random fluctuations. The overall modularity is thus defined as
$M\equiv {\text max}\{M^C, C=1,2,...\}$ and the corresponding
partition is assumed to be the most relevant.

Fig. 2 shows the analysis of the karate network as an example.
There we have considered $10^4$ random networks of the null model
generated with the rewiring technique \cite{random}. It can be
seen in Fig. 2(c) and (d) that the distribution of $Q_r^C$ is
perfect Gaussian, and hence can be well characterized by its
average ${\langle Q_r^C\rangle}$ and deviation $\delta_{Q_r^C}$.
Meanwhile ${\langle Q_r^C\rangle}$ is a function of $C$ (Fig.
2(a)); this is the reason why it is introduced as the denominator
in the definition of $M^C$. The results of $M^C$ (Fig. 2(b))
suggest that the partition of three communities ($C=3$; see Fig. 1
for the partition) is the most relevant. We have also studied the
dolphin network \cite{dolf1} and the most relevant partition
($C=2, M=0.6258$) is found to be exactly the same as the natural
split observed \cite{dolf2}. For another popular testing network
of the American college football teams \cite{ftb} our method
suggests the partition of 10 communities ($C=10, M=1.3345$).

\begin{figure}
\includegraphics[width=.9\columnwidth,clip]{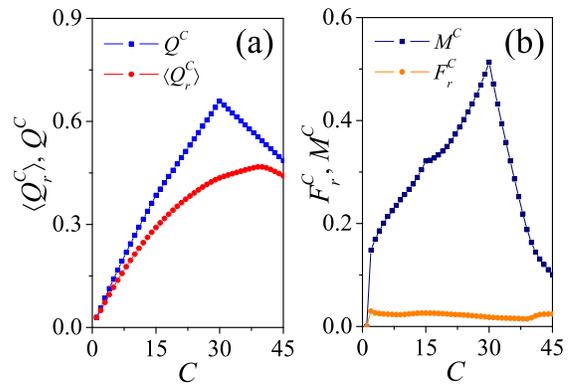}
\caption{(color online) The results of $Q^C$ and $\langle
Q_r^C\rangle$ (a), $M^C$ and $F_r^C$ (b) for a network of $30$
cliques on a circle. Each clique has $3$ nodes and the neighboring
cliques are connected with one link. The partition corresponding
to the largest $M^C$ ($C=30$) assigns each clique into a single
module accurately. }
\end{figure}

The fact that the distribution of $Q_r^C$ is Gaussian allows us to
define another useful quantity
\begin{eqnarray}
F_r^C=\frac{\delta_{Q_r^C}}{\langle Q_r^C\rangle} \label{eq-FC}
\end{eqnarray}
which gives how `modular' a random network (of the null model) can
be as a result of fluctuations. Obviously only the partitions of
the original network whose $M^C\gg F_r^C$ may suggest meaningful
community structures (see Fig. 2(b) for a comparison of $M^C$ and
$F_r^C$ in the karate network). This should be seen as a necessary
condition for the communities defined with the symmetric
modularity and it concludes our community detection scheme.

Now let us discuss two useful properties of the symmetric
modularity. First, {\it the community detection method based on it
has no resolution limit}. As an example \cite{reslmt} we consider
a network of $\cal{N}$ cliques sited on a circle. Each clique
contains $3$ nodes -- the smallest size for a meaningful module --
and any two neighboring cliques are linked with one edge. Our
scheme can identify all cliques (for ${\cal N}\geq 2$) without any
ambiguity (see Fig. 3 for ${\cal N}=30$ as an example). In this
simulation (and also in those for Fig. 4 and Fig. 5) $\langle
Q_r^C\rangle$ and $F_r^C$ are evaluated over $10^3$ random
networks with the same degree sequence. As a comparison, the
method with the asymmetric modularity $M_{NG}^C$ suggests instead
the partition of $10$ communities each containing $3$ neighboring
cliques ($M_{NG}=49/60$) due to its inherent resolution limit.
($M^C_{NG}=97/120$ and $43/60$ for $C=15$ and $C=30$ in this
case.)

This high resolution even makes our method applicable to the
hierarchical community networks -- a challenge for the quality
function method due to the multiple scales involved. In Fig. 4 we
present the partition results for the model hierarchical network
suggested in \cite{hier}: 256 nodes are divided into 16
compartments of equal size at the first level and every 4 of them
make a bigger compartment at the second level. The internal degree
of nodes at the first (second) level $z_{{\text {in}}_1}$
($z_{{\text {in}}_2}$) and the degree for the links between the
second level communities keep an average of $z_{{\text
{in}}_1}+z_{{\text {in}}_2}+z_{\text {out}}=18$ (hence the
hierarchical levels can be indicated by `$z_{{\text
{in}}_1}-z_{{\text {in}}_2}$'). We find that the hierarchical
structures are well characterized by the local maxima (also the
sharp turning points) on the $M^C$ curve indicating the relevant
scales ($C=4$ and $16$ in this case) and a higher level in
between. However, with the asymmetric modularity method
($M_{NG}^C$) no signal for the first level communities ($C=16$)
can be recognized.

\begin{figure}
\includegraphics[width=.9\columnwidth,clip]{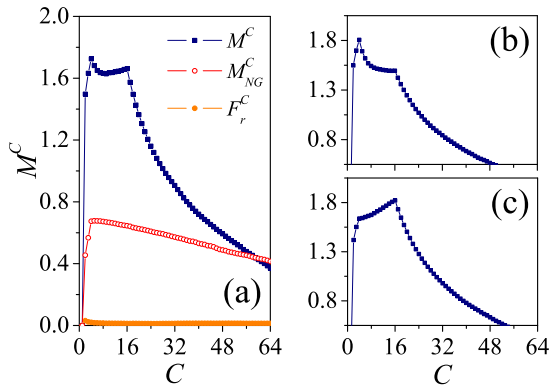}
 \caption{(color online) The symmetric modularity of
the hierarchial network \cite{hier} of type ($z_{{\text
{in}}_1}-z_{{\text {in}}_2}$) $14-3$ (a), $13-4$ (b) and $15-2$
(c). The relevant hierarchical scales $C=4$ and $16$ can be
related to the local maxima (also sharp turning points) on the
$M^C$ curve. The change of $M^C$ values at the maxima from (a) to
(c) reflects the competition of the two scales.}
\end{figure}

Second, {\it the symmetric modularity does not take large value
for a random network}. Careful studies of Erd\~os-R\'enyi (ER) and
Barabasi-Albertscale (BA) scale-free networks \cite{sf} are
summarized in Fig. 5. For an ER network with $N$ nodes and
connection probability $p$ studied there, we have verified that
$M^C$ is around zero and $|M^C|\sim F_r^C$ as implied by
definition. Meanwhile, the data suggest that $F_r^C$ may depend on
the degree sequence, but always takes the maximum value at $C=2$.
For this reason we have considered eight different degree
sequences for each $N$, $p$ pair and calculated their average and
the corresponding deviation (see Fig. 5(a)). It can be seen that
$\langle F_r^C\rangle$ is small and does not depend on $p$
significantly; more important as $N$ is increased it keeps
decreasing roughly in a power law $\sim N^{-0.75\pm 0.05}$. This
suggests that the community structure cannot be a {\it general}
property in ER networks. In addition, the dependence of $F_r^C$ on
the degree sequence is very weak ($\delta_{F_r^C}<0.05$),
suggesting the chance for finding meaningful community structure
in certain realizations of ER networks of {\it particular} degree
sequences is also very slim. The study of the scale-free networks
leads to the same results except that the power law dependence of
$\langle F_r^C \rangle$ on the network size is roughly $\sim
N^{-0.46\pm 0.04}$ instead.

In summary, we suggest the equivalence between the topological
information of a network and its complement should be considered
generally in the definition and detection of network structures.
As an important application we have focused on the community
partition problem and proposed a symmetric quality function. The
resulted community detecting scheme has a high resolution and can
be used to identify hierarchical community structures. In
addition, we have found that the effects of fluctuations on the
community structure are weak and decrease as the size of network
increases. This implies that the community structure is unlikely a
result of fluctuations when the size of the network is large
enough. The question of whether there are other relevant
symmetries and how they may provide insights into network
structures is interesting and deserves further efforts.

\begin{figure}
\includegraphics[width=.9\columnwidth,clip]{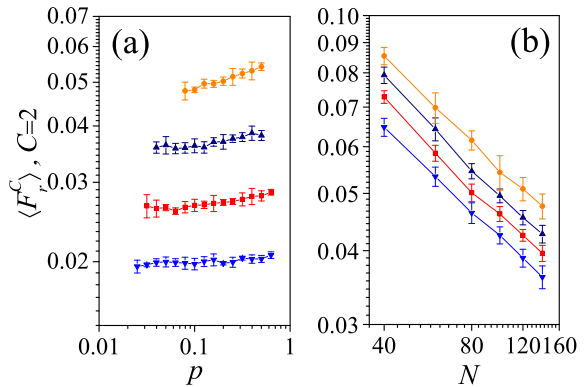}
\caption{(color online) The average of $F_r^C$ (see
Eq. (\ref{eq-FC})) and the corresponding deviation
$\delta_{F_r^C}$ (error bar) of the ER (a) and BA (b) networks
evaluated over eight different degree sequences. The four sets of
data from top to bottom correspond to $N=40, 60, 90$ and $135$ in
(a) and $m=5, 4, 3$ and $2$ in (b) with $m$ the defining parameter
of BA scale-free networks \cite{sf}.}
\end{figure}

This work is supported by Defense Science and Technology Agency
(DSTA) of Singapore under agreement POD0613356.

\end{document}